\documentclass[letterpaper, 10 pt, conference]{ieeeconf}

\IEEEoverridecommandlockouts

\usepackage[T1]{fontenc}

\usepackage[utf8]{inputenc}

\usepackage{mathtools}
\usepackage{amssymb}
\usepackage{bm}

\usepackage[center]{caption}
\usepackage{subfigure}

\usepackage{setspace}

\usepackage{wasysym}

\usepackage{hyperref}

\usepackage{tikz}
\usepackage{tikz-3dplot}
\usepackage{tkz-euclide}
\usetkzobj{all}
\usetikzlibrary{calc,decorations.markings,,intersections,shapes.misc}

\usepackage{graphicx}

\usepackage{multirow}

\usepackage{placeins}

\definecolor{dkgreen}{rgb}{0,0.6,0}
\definecolor{light-gray1}{gray}{0.40}
\definecolor{light-gray2}{gray}{0.60}
\definecolor{light-gray3}{gray}{0.75}

\newcommand*{\Scale}[2][4]{\scalebox{#1}{$#2$}}%

\newcommand{\sss}[1]{\scriptscriptstyle {#1}}

\newcommand{\tp}{^{\sss{T}}}

\newcommand{\twocol}[2]
 {\begin{bmatrix}
    {#1} \\
    {#2}
  \end{bmatrix}}

\newcommand{\tworow}[2]
 {\begin{bmatrix} {#1}&{#2} \end{bmatrix}}

\newcommand{\Rn}[1][n]{\mathbb{R}^{\sss{#1}}}

\newcommand{\sk}[1]{\mathcal{S}\left({#1}\right)}
\newcommand{\OP}[1]{\Pi\left({#1}\right)}

\newtheorem{prob}{Problem}

\newcommand{\emb}{\mathbf{e}}
\newcommand{\fmb}{\mathbf{f}}
\newcommand{\gmb}{\mathbf{g}}

\newcommand{\nmb}{\mathbf{n}}

\renewcommand{\pmb}{\mathbf{p}}

\newcommand{\rmb}{\mathbf{r}}

\newcommand{\umb}{\mathbf{u}}
\newcommand{\vmb}{\mathbf{v}}

\newcommand{\xmb}{\mathbf{x}}
\newcommand{\ymb}{\mathbf{y}}

\newcommand{\Tmb}{\mathbf{T}}

\newcommand{\Rmat}{\ensuremath{\mathcal{R}}} 	%rotation matrix
\newcommand{\zvec}{\mathbf{0}} 				%zeros vector
\newcommand{\SO}[1][2]{\mathcal{SO}(#1)}

\newcommand{\embThree}{\emb_{\sss{3}}}

\newcommand{\pQ}{\pmb}
\newcommand{\pQDes}{\pmb^{\sss{d}}}
\newcommand{\vQ}{\vmb}
\newcommand{\vQDes}{\vmb^{\sss{d}}}
\newcommand{\pQDot}{\dot{\pmb}}
\newcommand{\pQDDot}{\ddot{\pmb}}
\newcommand{\pQDesDDot}{\ddot{\pmb}^{\sss{d}}}
\newcommand{\pL}{\pmb_{\sss{L}}}
\newcommand{\vL}{\vmb_{\sss{L}}}

\newcommand{\pLstar}{\pmb_{\sss{L}}^{\sss{d}}}

\newcommand{\mQ}{m}
\newcommand{\RQ}{\Rmat}
\newcommand{\RQDot}{\dot{\Rmat}}
\newcommand{\rQ}{\rmb_{\sss{3}}}
\newcommand{\rQTP}{\rmb_{\sss{3}}\tp}

\newcommand{\rQDot}{\dot{\rmb}_{\sss{3}}}

\newcommand{\wQ}{\bm{\omega}}
\newcommand{\wQDot}{\dot{\bm{\omega}}}
\newcommand{\wThree}{\bm{\omega}_{\sss{3}}}
\newcommand{\wThreeDot}{\dot{\bm{\omega}}_{\sss{3}}}

\newcommand{\Tau}{\bm{\tau}_{\sss{m}}}

\newcommand{\TauQ}{\bm{\tau}}
\newcommand{\TauThree}{\bm{\alpha}}

\usepackage{epstopdf}

\begin{document}

	\title{Quadrotor Controller}
	
	\date{\today}
	\author{
		Pedro~O.~Pereira and Dimos~V.~Dimarogonas
		% P.P., D.D.
		\thanks{ %
			The authors are with the School of Electrical Engineering, KTH Royal Institute of Technology, SE-100 44, Stockholm, Sweden. %
			\texttt{\{ppereira, boskos, dimos\}@kth.se}. This work was supported by funding from the European Union's Horizon 2020 Research and Innovation Programme under the Grant Agreement No.644128 (AEROWORKS).
		}
	}
	
	\maketitle	

	\begin{abstract}
		In this paper we construct a trajectory tracking controller for a quadrotor system by finding a coordinate change which transforms the quadrotor's vector field into that of a thust propelled system.
In a thrust propelled system, the goal is to stabilize its position around the origin, while the system is actuated by a one dimensional acceleration/thrust along a direction vector, by a time-varying gravity, and by the angular acceleration of the direction vector.
For this system, a solution has been proposed in~\cite{ecc2016} based on the implicit knowledge of a bounded controller for a double integrator system, and on the implicit knowledge of a Lyapunov function that guarantees the origin is asymptotically stable for the double integrator controlled by the bounded controller. 
In this paper, we present two alternative bounded controllers for a double integrator system, and corresponding Lyapunov functions.
	\end{abstract}

	%\linenumbers

	\section{Introduction}
	Many controllers for trajectory tracking of aerial vehicles have been proposed.
Controllers based on a linearized model around the hover condition have been successfully deployed, but their applicability is limited to trajectories where the flight envelop remains in small angles ~\cite{hoffmann2008quadrotor,koo1998output}.
Controllers based on an inner and outer loop strategy have also been successfully applied, where the inner loop is responsible for attitude control loop, while the outer loop is responsible for the position control~\cite{michael2010grasp}.
The quadrotor dynamics depend on the vehicle's rotation matrix, most control strategies also provide a control law for the space corresponding to the yaw motion.
Different parameterizations for the vehicle's rotation matrix  have also been used, such as euler angles~\cite{michael2010grasp}, and unit quaternions~\cite{roberts2011adaptive, casau2013global}.
Controllers that guarantee trajectory tracking for all initial conditions, can also be found~\cite{casau2013global}.

	\section{Modeling}
\begin{figure}
	\centering
	\includegraphics[clip=true,trim=0cm 0cm 0cm 0cm,width=0.4\textwidth]
	{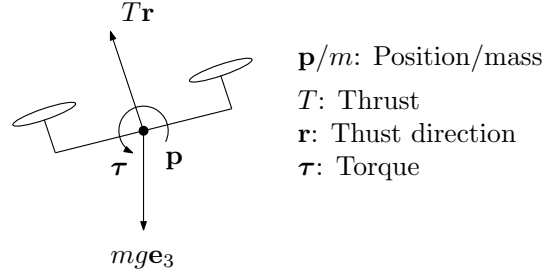}		
	\caption{Modeling for quadrotor transporting load with a rigid manipulator}
	\label{fig:Modeling}
\end{figure}

	A sketch of a simplified model of a quadrotor is presented in Fig.~\ref{fig:Modeling}.

We denote $\pQ\in \Rn[3]$ as the quadrotor's positions, $\vQ(t) := \pQDot(t)$  as the quadrotor's velocity, and $\mQ\in \Rn[{}]_{\sss{>0}}$ as the quadrotor's mass.
Also we denote $\RQ\in\SO[3]$ as the quadrotor's rotation matrix, $\rQ := \RQ \embThree \in  \mathcal{S}^{\sss{2}}$ as the quadrotor's direction where thrust is provided, and $J = J\tp \in \Rn[3\times 3]$ as the quadrotor matrix of inertia.
Finally, we denote $T \in \Rn[{}]$ as the quadrotor's thrust, $\TauQ \in \Rn[3]$ as the quadrotor's torque,  where $T$, $\TauQ$ and $\Tau$ are all assumed to be a control inputs. 

The system quadrotor and load is evolves according with the dynamics
\begin{align}
	\mQ \pQDDot(t) 
	& =
	k_{\sss{T}} T(t) \rQ(t) 
	-
	\mQ g \embThree
	\label{eq:f1}
	\\
	\RQDot(t) 
	&=
	\RQ(t)\sk{\wQ(t)},
	\label{eq:f2}
	\\
	J \wQDot(t)
	& =
	- \sk{\wQ(t)} J \wQ(t)
	+
	\TauQ(t)
	\label{eq:f3}
\end{align}
where $k_{\sss{T}}  \in \Rn[{}]_{> 0}$ is a constant gain.

\begin{prob}
	\label{prob:Problem}
	Given the system~\eqref{eq:f1}-\eqref{eq:f3}, with known $k_{\sss{T}}$, and a desired load trajectory $\pQDes(\cdot) \in \mathcal{C}^{\sss{4}}(\Rn[{}]_{\sss{\ge 0}})$, design $T: \Rn[{}]_{\sss{\ge 0}} \mapsto \Rn[]$ and $\TauQ : \Rn[{}]_{\sss{\ge 0}} \mapsto \Rn[3]$ such that $\lim_{\sss{t \rightarrow \infty }} (\pL(t) - \pLstar(t)) = \zvec$. 
\end{prob}

%\begin{prob}
%	\label{prob:ProblemDisturbed}
%	Given the system~\eqref{eq:f1}-\eqref{eq:f3}, with unknown $k_{\sss{T}}$, and a desired load trajectory $\pQDes(\cdot) \in \mathcal{C}^{\sss{4}}(\Rn[{}]_{\sss{\ge 0}})$, design $T: \Rn[{}]_{\sss{\ge 0}} \mapsto \Rn[]$ and $\TauQ : \Rn[{}]_{\sss{\ge 0}} \mapsto \Rn[3]$ such that $\lim_{\sss{t \rightarrow \infty }} (\pL(t) - \pLstar(t)) = \zvec$ and $\lim_{\sss{t \rightarrow \infty }} (\nmb(t) - \nmb^{\sss{d}}(t)) = \zvec$. 
%\end{prob}

	\section{Known $k_{\sss{T}}$}
	It is easy to verify that the system~\eqref{eq:f1}-\eqref{eq:f2} is flat with respect to the output  $\ymb = \pmb(t)$.

Let us also now define the variables
\begin{align}
	&\pmb(t) := \pQ(t) - \pQDes(t)
	\label{eq:Position}
	\\
	&\vmb(t) := \vQ(t) - \vQDes(t)	
	\label{eq:Velocity}
	\\
	&
	\gmb(t)
	:= 
	g\embThree + \pQDesDDot(t),
	\\
	&
	\rQ(t)
	:= 
	\RQ(t) \embThree,	
	\\
	&
	\wThree(t)
	:= 
	\OP{\rQ(t)} \RQ(t)\wQ(t),
	\label{eq:Omega3}
\end{align} 
where we emphasize that, by construction, $\wThree\tp(t) \rQ\tp(t) = 0$ for all $t \in \Rn[{}]$.
We also denote
\begin{align}
	\xmb\tp
	:=	
	\begin{bmatrix}
		\bar{\xmb}\tp
		& 
		\wThree\tp
	\end{bmatrix}	
	:=		
	\begin{bmatrix}
		[\pmb\tp
		&
		\vmb\tp
		&
		\rQTP
		]
		&
		\wThree\tp
	\end{bmatrix}.
	\label{eq:stateQuad}
\end{align}
The goal of Problem~\ref{prob:Problem} is then equivalent to $\lim_{\sss{t \rightarrow \infty}} \pmb(t) = \zvec$ and $\lim_{\sss{t \rightarrow \infty }} (\nmb(t) - \nmb^{\sss{d}}(t)) = \zvec$.

For reasons that will be apparent later, we choose $T(t)$ and $\TauQ(t)$ as
\begin{align}
	T(t)
	=
	& 
	\frac{\mQ}{k_{\sss{T}}}
	a(t)
	\label{eq:ThrustQuad}
	\\
	\TauQ(t) 
	=
	& 
	\sk{\wQ(t)} J \wQ(t)
	+ 
	J \embThree \tau_{\sss{3}}(t)
	\\
	&
	J
	\sk{\embThree} 
	\left(
		\RQ\tp(t)
		\TauThree(t)
		-
		\wQ(t) (\embThree\tp \wQ(t))
	\right),
	\label{eq:TauQuad}
\end{align}
where $a(t)$ and $\TauThree(t)$ are to be designed later (and $\tau_{\sss{3}}(t)$ is a degree of freedom that can be used to accomplish other goals).
From~\eqref{eq:TauQuad}, it follows that ~\eqref{eq:f3} becomes
\begin{align}
	\wQDot(t) = \sk{\embThree} (\RQ\tp(t) \TauThree(t)  - \wQ(t) (\embThree\tp \wQ(t))) + \embThree \tau_{\sss{3}}(t).
	\label{eq:OmegaDot2}
\end{align}
Also, note that $\rQDot(t) = \sk{\RQ(t) \wQ(t)} \rQ(t) = \sk{\OP{\rQ(t)} \RQ(t) \wQ(t)} \rQ(t) = \sk{\wThree(t)} \rQ(t)$ and therefore (for brevity, we omit the time dependencies below)
\begin{align}
		\wThreeDot
		= & 
		\OP{\rQ} (\RQDot\wQ + \RQ \wQDot)
		\\
		&
		-
		\left(\rQDot\rQTP + \rQ\rQDot\tp \right)\RQ \wQ
		\\
		\overset{\sss{\eqref{eq:OmegaDot2}}}{=} 
		&
		\OP{\rQ} (\sk{\rQ} ( \TauThree  - \RQ\wQ (\embThree\tp\wQ)) + \rQ \tau_{\sss{3}})
		\\
		&
		-
		\rQDot(\rQTP\RQ \wQ)
		-
		\rQ \rQDot\tp \RQ \wQ
		\\
		= 
		&
		\sk{\rQ} (\TauThree - \RQ \wQ (\rQTP\RQ \wQ)
		\\
		&
		-
		\sk{\wThree} \rQ (\rQTP\RQ \wQ)
		\\
		= &
		\sk{\rQ} \TauThree
		\label{eq:OmegaDotAux2}
\end{align}

If~\eqref{eq:ThrustQuad} and~\eqref{eq:TauQuad} are chosen, then 
\begin{align}
	&\dot{\pmb}(t) \overset{\sss{\eqref{eq:Position},\eqref{eq:Velocity}}}{=} \vL(t)
	\\
	&\dot{\vmb}(t) 
	\overset{\sss{\eqref{eq:f1},\eqref{eq:ThrustQuad}}} =   
	a(t)\nmb(t) 
	-
	\gmb(t)
	\\
	&
	\rQDot(t)
	\overset{\sss{\eqref{eq:f3},\eqref{eq:Omega3}}}{=}
	\sk{\wQ(t)}
	\rQ(t)
	\\
	&
	\wThreeDot(t)
	\overset{\sss{\eqref{eq:OmegaDotAux2}}}{=} 
	\sk{\nmb(t)} \TauThree(t),
\end{align}

We can then write the dynamics of $\xmb(t)$  as $\dot{\xmb}(t) = \fmb(t,\xmb(t),a(t),\TauThree(t))$, where
\begin{align}		
	\fmb(t,\xmb,a,\TauThree)
	& :=
	\begin{bmatrix}
			\vmb
			\\
			a \rQ - \gmb(t)
			\\
			\sk{\wThree}\rQ
			\\
			\sk{\rQ} \TauThree
	\end{bmatrix}
	\label{eq:VectorFieldCM}	
\end{align}

The vector field $\fmb_{\sss{u}}(t,\xmb,a,\TauThree)$ is that of a vector thrusted system, and a general solution that guarantees $\lim_{\sss{t \rightarrow \infty}} \pmb(t) = \zvec$ for $\dot{\xmb}_{\sss{1}}(t) = \fmb_{\sss{u}}(t,\xmb(t),a(t),\TauThree(t))$ can be found in~\cite{}.

In~\cite{}, we can find functions $T_{\sss{cl}}(t,\bar{\xmb})$  and $\bm{\tau}_{\sss{cl}}(t,\xmb)$ such that if  $\dot{\xmb}(t) = \fmb_{\sss{u}}^{\sss{cl}}(t,\xmb(t))$ where
\begin{align}
	\fmb_{\sss{u}}^{\sss{cl}}(t,\xmb) := \fmb_{\sss{u}}(t,\xmb,T_{\sss{cl}}(t,\bar{\xmb}),\bm{\tau}_{\sss{cl}}(t,\xmb))
	\label{eq:fClosedLoopUniThrust}
\end{align}
it follows that $\lim_{\sss{t \rightarrow \infty}} \pmb(t) = \zvec$.
Also, for 
\begin{align}
	V_{\sss{3}}(t,\xmb)
	=
	V_{\sss{1}}(\pmb,\vmb)
	+
	V_{\sss{\theta}}(\xi(t,\bar{\xmb}))
	+
	V_{\sss{\omega}}(\emb_{\sss{\omega}}(t,\xmb)),
	\label{eq:V3}
\end{align}
the following is satisfied
\begin{align}
	 & \frac{\partial V_{\sss{3}}(t,\xmb)}{\partial t}
	+
	\frac{\partial V_{\sss{3}}(t,\xmb)}{\partial \xmb}
	\fmb_{\sss{x}}^{\sss{cl}}(t,\xmb(t))
	=
	\\
	= &
	-
	W_{\sss{1}}(\pmb,\vmb)
	-
	2
	k_{\sss{\omega}}
    V_{\sss{\omega}}(\emb_{\sss{\omega}}(t,\xmb))
    \\
    &
   	-
   	k_{\sss{\theta}}
   	V'_{\sss{\theta}}(\xi(t,\bar{\xmb}))
   	\xi(t,\bar{\xmb}) (2 - \xi(t,\bar{\xmb}))
	\\
	=: &
	- W_{\sss{3}}(t,\xmb) \le 0,
	\label{eq:W3}
\end{align}
and, along a trajectory $\xmb(t)$ with $\dot{\xmb}(t) = \fmb_{\sss{u}}^{\sss{cl}}(t,\xmb(t))$, $\dot{V}_{\sss{3}}(t,\xmb(t)) = -W_{\sss{3}}(t,\xmb(t)) \le 0$.

As such, if we choose $a(t) = T_{\sss{cl}}(t,\bar{\xmb}(t))$  and $\TauThree(t) = \bm{\tau}_{\sss{cl}}(t,\xmb(t))$, we can guarantee that $\lim_{\sss{t \rightarrow \infty}} \pmb(t) = \zvec \Rightarrow \lim_{\sss{t \rightarrow \infty}} (\pQ(t) -\pQDes(t)) = \zvec$.

%	\section{Known $k_{\sss{T}}$}
%	\input{./Contents/FullyActuated}

	\section{Double Integrator}
	The control laws $T_{\sss{cl}}(t,\bar{\xmb})$ and $\bm{\tau}_{\sss{cl}}(t,\xmb)$ in \cite{ecc2016} have been constructed based on the implicit knowledge of a Lyapunov function $V_{\sss{1}}(\cdot,\cdot): \Rn[] \times \Rn[] \mapsto \Rn[{}]_{\ge 0}$, such that for the double integrator vector field $[v \, u(p,v))]\tp$, it holds that
\begin{align}
	\frac{\partial V(p,v)}{\partial p} v
	+
	\frac{\partial V(p,v)}{\partial p} u(p,v)
	= :
	- W(p,v)
	\le 0.
\end{align}
and that $|u(p,v)| \le u^{\sss{\infty}}$ for all $(p,v) \in \Rn[{}]\times \Rn[{}]$.

Here, we present two options, i.e., two pairs $(u(\cdot,\cdot), V_{\sss{1}}(\cdot,\cdot))$ that satisfy the previous two conditions.

\subsection{Option 1}
Consider $\sigma \in \Sigma$ and $\varrho \in \tilde{\Sigma}$, and a symmetric function $\xi: \Rn[] \mapsto \Rn[]$ that satisfies
\begin{align}
	& \xi \in \mathcal{C}^{\sss{3}}(\Rn[{}])
	\label{eq:XiProperty1}
	\\
	& |\xi(s)| \ge |s| \wedge \xi^{\sss{\prime}}(s)> 0, \forall s \in \Rn[{}]
	\label{eq:XiProperty2}
	\\
	& \xi(s) = s, \, \forall |s| \le \bar{\sigma}
	\label{eq:XiProperty3}
	\\
	& \sup_{\sss{s \in \Rn[{}]}} \left|\frac{s}{\xi'(s)}\right| =: M < \infty
	\label{eq:XiProperty4}
\end{align}
where $M \ge \bar{\sigma}$ holds necessarily, owing to~\eqref{eq:XiProperty3}. From the second condition, it follows that
\begin{align}
	&
	\frac{v + \sigma(p)}{\xi(v) + \sigma(p)}
	\overset{\sss{\eqref{eq:XiProperty3}}}{=} 1,
	\, \forall 
	(p,v) \in \Rn[{}] \times [-\bar{\sigma},\bar{\sigma}]
	\\
	&
	\frac{v + \sigma(p)}{\xi(v) + \sigma(p)}
	\overset{\sss{\eqref{eq:XiProperty2}}}{\le} 1,
	\, \forall 
	(p,v) \in \Rn[{}] \times \Rn[{}]	
\end{align}
and
\begin{align}
	\sup_{\sss{s \in \Rn[{}]}} \frac{1}{|\xi'(s)|} \le \frac{M}{\bar{\sigma}},
\end{align}
since $\xi'(s) = 1$ for $|s| \le \bar{\sigma}$, and $\frac{1}{\xi'(s)} \le \frac{M}{s}$ for all $s \in \Rn[{}]$.

Consider then $k_{\sss{\sigma}} > 0$, and the control law
\begin{align}
	u(p,v)
	=
	&
	-\varrho(\xi(v) + \sigma(p))
	-
	\frac{v + \sigma(p)}{\xi(v) + \sigma(p)}
	\frac{k_{\sss{\sigma}} \sigma(p)}{\xi^{\sss{\prime}}(v)}
	-
	\\
	&
	-
	\sigma^{\sss{\prime}}(p)\frac{v}{\xi^{\sss{\prime}}(v)}
	\label{eq:ControllerOption1}
\end{align}
is then bounded, namely
\begin{align}
	\sup_{\sss{(p,v) \in \Rn[2]}} |u(p,v)| \le M (k_{\sss{\sigma}} + \bar{\sigma}^{\sss{\prime}})  + \bar{\varrho} < \infty.
\end{align}

Consider the function
\begin{align}
	V(p,v)
	=
	k_{\sss{\sigma}}
	\int_{\sss{0}}^{\sss{p}}
	\sigma(s) ds
	+
	\frac{1}{2}
	(\xi(v) + \sigma(p))^2
\end{align}
for which it follows that
\begin{align}
	& \frac{\partial V(p,v)}{\partial p} v
	+
	\frac{\partial V(p,v)}{\partial p} u(p,v)
	= \\
	= &
	k_{\sss{\sigma}} \sigma(p) v
	+
	(\xi(v) + \sigma(p))
	\left(
		\xi^{\sss{\prime}}(v) u(p,v)
		+ 
		\sigma^{\sss{\prime}}(p) v
	\right)
	= 
	\\
	\overset{\sss{\eqref{eq:ControllerOption1}}}{=} &
	k_{\sss{\sigma}} \sigma(p) v
	-
	(v + \sigma(p))
	k_{\sss{\sigma}} \sigma(p)
	\\
	&
	-
	\xi^{\sss{\prime}}(v)
	(\xi(v) + \sigma(p))
	\varrho(\xi(v) + \sigma(p))
	\\
	= &
	-
	k_{\sss{\sigma}} \sigma^2(p)
	-
	\xi^{\sss{\prime}}(v)
	(\xi(v) + \sigma(p))
	\varrho(\xi(v) + \sigma(p))
	\\
	=: &
	-
	W(p,v) \le 0
\end{align}

Notice that if we pick $\varrho(s) = 0$, it follows that $W(p,v) = \sigma^2(p)$. 
For this scenario, it still follows that $\lim_{\sss{t \rightarrow \infty}} W(p(t),v(t)) \Leftrightarrow \lim_{\sss{t \rightarrow \infty}}p(t) = 0$.

\subsection{Option 2}

Denote $\Sigma$ as the family of functions $\chi:\Rn[] \mapsto \Rn[]$ that satisfy (1) $\chi(s) s > 0$ for $s \ne 0$, (2) $\sup_{\sss{s \in \Rn[{}]}}|\chi(s)| \le \chi_{\sss{\infty}} < \infty$, (3) $\chi \in \mathcal{C}^{\sss{3}}(\Rn[{}])$, (4) $\sup_{\sss{s \in \Rn[{}]}} |\chi^{\sss{\prime}}(s)| \le \bar{\chi}^{\sss{\prime}} < \infty$. 
As such, for $\chi \in \Sigma$, it follows that
\begin{align}
	\frac{\chi(r)}{\bar{\chi}^{\sss{\prime}} r}
	,
	\frac{\chi^{\sss{\prime}}(r)}{\bar{\chi}^{\sss{\prime}}} 
	\in [-1,1],
	\forall r \in \Rn[{}]
	\label{eq:SigmaFamily}
\end{align}
and therefore, since (1) holds, it also follows that
\begin{align}
	&
	\left(
		1 - \frac{\chi^{\sss{\prime}}(r)}{\bar{\chi}^{\sss{\prime}}}
	\right)
	\chi(r) r \overset{\sss{\eqref{eq:SigmaFamily}}}{\ge} 0
	\Rightarrow
	\\
	\Rightarrow 
	&
	\int_{\sss{0}}^{s} \chi(r) dr
	\ge
	\int_{\sss{0}}^{s} \frac{\chi^{\sss{\prime}}(r)}{\bar{\chi}^{\sss{\prime}}}\chi(r) dr
	=
	\frac{1}{2 \bar{\chi}^{\sss{\prime}}}
	\int_{\sss{0}}^{s} \frac{d \bar{\chi}^{2}(r)}{dr}  dr
	\\
	\Leftrightarrow &
	\int_{\sss{0}}^{s} \chi(r) dr
	\ge 
	\frac{\bar{\chi}^{2}(s)}{2 \bar{\chi}^{\sss{\prime}}}.
	\label{eq:SigmaFamilyProp1}
\end{align}
and that
\begin{align}
	&
	\left(
		1 
		- 
		\frac{\chi^{\sss{\prime}}(r)}{\bar{\chi}^{\sss{\prime}}}
		\frac{\chi(r)}{\bar{\chi}^{\sss{\prime}} r}
	\right)
	\bar{\chi}^{\sss{\prime 2}} r r
	\overset{\sss{\eqref{eq:SigmaFamily}}}{\ge} 
	0
	\Rightarrow
	\\
	\Rightarrow &
	\int_{\sss{0}}^{s} \bar{\chi}^{\sss{\prime 2}} r - \chi^{\sss{\prime}}(r) \chi(r) dr
	\ge 
	0.
	\label{eq:SigmaFamilyProp2}
\end{align}
We also denote $\tilde{\Sigma} \subset \Sigma$ as the family of functions $\chi:\Rn[] \mapsto \Rn[]$ that satisfy (1) $\chi \in \Sigma$, and (2) $\chi^{\sss{\prime}}(s) > 0 $ for all $s \ne 0$.

Consider the control law
\begin{align}
	u(p,v)
	=
	-
	\sigma(p)
	-
	\varrho(v)
	\label{eq:ControllerOption2}
\end{align}
where $\sigma \in \Sigma$ and $\varrho \in \tilde{\Sigma}$.
Therefore, 
\begin{align}
	\sup_{\sss{(p,v) \in \Rn[2]}} |u(p,v)| \le \bar{\sigma}  + \bar{\varrho} < \infty.
\end{align}

Given $\beta \in (0,1)$, consider then the function
\begin{align}
	V_{\sss{4}}(p,v) = V_{\sss{1}}(p) + V_{\sss{2}}(v) + V_{\sss{3}}(p,v),
	\label{eq:LyapunovStability}
\end{align}
where (denote $\frac{\partial V_{\sss{i}}}{\partial p} v + \frac{\partial V_{\sss{i}}}{\partial v} u(p,v) =: - W_{\sss{i}}$ for $i = \{1, 2,3,4\}$)
\begin{align}
	V_{\sss{1}}(p) &= \bar{\sigma}^{\sss{\prime}} \int_{0}^{p} \sigma(s) ds > 0,
	\label{eq:V1}
	\\
	W_{\sss{1}}(p,v) &=  -\bar{\sigma}^{\sss{\prime}} \sigma(p) v
	\label{eq:W1}
\end{align}
and
\begin{align}
	V_{\sss{2}} 
	= &
	\beta \int_{0}^{v} (\bar{\varrho}^{\sss{\prime 2}} s - \varrho^{\sss{\prime}}(s)\varrho(s)) ds 
	\overset{\sss{\eqref{eq:SigmaFamilyProp2}}}{\ge} 
	0,
	\\
	W_{\sss{2}}(p,v) 
	= &  
	\beta(\bar{\varrho}^{\sss{\prime 2}}v - \varrho^{\sss{\prime}}(s)\varrho(v)) \sigma(p) +
	\\
	&
	\beta(\bar{\varrho}^{\sss{\prime 2}}v - \varrho^{\sss{\prime}}(s)\varrho(v))  \varrho(v)
	\\
	:=
	& 
	W_{\sss{2,1}}(p,v)+ 
	W_{\sss{2 2}}(v)
	\label{eq:W2}	
\end{align}
and
\begin{align}
	V_{\sss{3}}(p,v) 
	= 
	&
	\beta \bar{\varrho}^{\sss{\prime 2}} \int_{0}^{p} \sigma(s) ds 
	+ 
	\beta \sigma(p) \varrho(v)
	+ 
	\bar{\sigma}^{\sss{\prime}}\frac{v^{2}}{2},
	\label{eq:V3}
	\\
	\overset{\sss{\eqref{eq:SigmaFamilyProp2}}}{\ge}
	& 
	\frac{1}{2}
	\tworow{\sigma(p)}{v}
	\begin{bmatrix}
		\frac{\beta \bar{\varrho}^{\sss{\prime 2}}}{\bar{\sigma}^{\sss{\prime}}} & \frac{\beta \varrho(v)}{v} \\
		\frac{\beta \varrho(v)}{v} & \bar{\sigma}^{\sss{\prime}}
	\end{bmatrix}
	\twocol{\sigma(p)}{v},
	\label{eq:MatrixLyap}
	\\
	W_{\sss{3}}(p,v)
	= & 
	-
	\beta \bar{\varrho}^{\sss{\prime 2}} \sigma(p) v
	-
	\beta \sigma^{\sss{\prime}}(p) \varrho(v) v
	+
	\\
	&
	\beta \sigma^2(p) \varrho^{\sss{\prime}}(v)
	+
	\beta \sigma(p) \varrho^{\sss{\prime}}(v)	\varrho(v) 
	+
	\\
	&
	\bar{\sigma}^{\sss{\prime}}
	v (\sigma(p) + \varrho(v))
	\\
	= &
	-
	W_{\sss{1}}(p)
	-
	W_{\sss{2,1}}(p,v) 
	+
	\beta \varrho^{\sss{\prime}}(v) \sigma^2(p)	
	\\
	& 
	\bar{\sigma}^{\sss{\prime}} v\varrho(v)
	\left(
		1 - \beta \frac{\sigma^{\sss{\prime}}(p)}{\bar{\sigma}^{\sss{\prime}}}
	\right)
	>0	
	\label{eq:W3}
\end{align}
The latter inequality in~\eqref{eq:MatrixLyap} follows since the matrix in~\eqref{eq:MatrixLyap} has positive diagonal elements and since its  determinant is also positive, namely
\begin{equation}
	\beta \bar{\varrho}^{\sss{\prime2}} \left(1 - \beta \left(\frac{\varrho(v)}{\bar{\varrho}^{\sss{\prime}} v}\right)^2\right)
	\overset{\substack{\varrho \in \tilde{\Sigma}\\ \beta\in(0,1)}}{>}
	0
	,
\end{equation}
which means $V_{\sss{3}}(p,v) > 0$ for any $\beta \in (0\,\,1)$.
Combining~\eqref{eq:W1}, \eqref{eq:W2} and~\eqref{eq:W3}, it follows that
\begin{align}
	W_{\sss{4}}(p,v) = 
	&
	\beta \varrho^{\sss{\prime}}(v) \sigma^2(p)
	+
	W_{\sss{2 2}}(v)
	+
	\\
	&
	\bar{\sigma}^{\sss{\prime}} v\varrho(v)
	\left(
		1 - \beta \frac{\sigma^{\sss{\prime}}(p)}{\bar{\sigma}^{\sss{\prime}}}
	\right)	
	\overset{\substack{\varrho \in \tilde{\Sigma}\\ \beta\in(0,1)}}{>}
	0
\end{align}
and note that ($\Rn[\star] = \Rn[{}]\cup \pm \infty$)
\begin{align}
	(0,0)
	=
	\left\{
		(\bar{p},\bar{v})
		\in
		\Rn[\star]
		\times
		\Rn[\star]
		:
		\lim_{\sss{(p,v) \rightarrow (\bar{p},\bar{v})}} W_{\sss{4}}(p,v) = 0	
	\right\}
\end{align}	
which means that if $\lim_{\sss{t \rightarrow \infty}} W_{\sss{4}}(p(t),v(t)) \Leftrightarrow \lim_{\sss{t \rightarrow \infty}}p(t) = 0$.

%	\section{Simulations}
%	\input{./Contents/Simulations}
		
	%%%% Bibliografia %%%%
	\bibliographystyle{unsrt} %agsm
	\bibliography{bibliography}
    % \clearpage
%		
	\appendices
%	\section{...}
%	\input{./Contents/AppendixA}
	
%	\section{...}
	Given $\umb(\pmb,\vmb)$ and $V_{\sss{1}}(\pmb,\vmb)$ (see~\cite{ecc2016}):
\begin{align}
	\Tmb^{\sss{\star}}(t,\pmb,\vmb) & := \gmb(t) + \umb(\pmb,\vmb),
	\label{eq:TStar}
	\\
	\nmb^{\sss{\star}}(t,\pmb,\vmb) 
	& 
	:=
	\frac{\Tmb^{\sss{\star}}(t,\pmb,\vmb)}{\|\Tmb^{\sss{\star}}(t,\pmb,\vmb)\|},
	\label{eq:nStar}
	\\
	\xi(t,\bar{\xmb})
	& := 
	1 - \nmb\tp \nmb^{\sss{\star}}(t,\pmb,\vmb) ,
	\\
	\Scale[0.95]{
		\bm{\omega}^{\sss{\star}}(t,\bar{\xmb})
	}
	& 
	\Scale[0.95]{
		: =
		\sk{\nmb^{\sss{\star}}(t,\pmb,\vmb)}
		\frac{
			\frac{\partial \Tmb^{\sss{\star}}(t,\pmb,\vmb)}{\partial t}
			+
			[
				\frac{\partial\Tmb^{\sss{\star,T}}}{\partial \pmb} 
				\,
				\frac{\partial\Tmb^{\sss{\star,T}}}{\partial \vmb} 
			]
			\fmb_{\sss{pv}}^{\sss{cl}}(t,\bar{\xmb})
		}
		{\|\Tmb^{\sss{\star}}(t,\pmb,\vmb)\|}
	},
	\label{eq:OmegaStar}
	\\		
	\bm{\omega}^{\sss{d}}(t,\bar{\xmb})
	& :=
    \bm{\omega}^{\sss{\star}}(t,\bar{\xmb})
	-
	k_{\sss{\theta}} \sk{\nmb^{\sss{\star}}(t,\pmb,\vmb)}\nmb
	\\
	&
	\hphantom{:=}
	-
	\frac{\|\Tmb^{\sss{\star}}(t,\pmb,\vmb)\|}{V_{\sss{\theta}}'(\xi(t,\bar{\xmb}))} 
	\sk{\nmb}\frac{\partial V_{\sss{1}}(\pmb,\vmb)}{\partial \vmb},
	\label{eq:OmegaD}	
	\\
	\emb_{\sss{\omega}}(t,\xmb) 
	&
	:= \sk{\nmb}(\bm{\omega} - \bm{\omega}^{\sss{d}}(t,\bar{\xmb})),
	\\
	\bm{\tau}^{\sss{d}}(t,\xmb)
	& :=
	\frac{\partial \bm{\omega}^{\sss{d}}(t,\bar{\xmb})}{\partial t}
	+
	\frac{\partial \bm{\omega}^{\sss{d,T}}(t,\bar{\xmb})}{\partial \bar{\xmb}}
	\tilde{\fmb}_{\sss{\bar{x}}}(t,\xmb)
	,	
	\label{eq:TauD}	
	\\
	T_{\sss{cl}}(t,\bar{\xmb})
	& :=
	\nmb\tp \Tmb^{\sss{\star}}(t,\pmb,\vmb) ,
	\label{eq:ThrustClosedLoop}		
	\\
	\bm{\tau}_{\sss{cl}}(t,\xmb)
	&
	:=
	\sk{\nmb}
	\bm{\tau}^{\sss{d}}(t,\xmb)
	+
	\OP{\nmb}
	\bm{\omega}^{\sss{d}}(t,\bar{\xmb})
	\nmb\tp\bm{\omega}^{\sss{d}}(t,\bar{\xmb})
	\\
	&
	\hphantom{:=}
	-	
	k_{\sss{\omega}} \emb_{\sss{\omega}}(t,\xmb)
	-
	\frac{V_{\sss{\theta}}'(\cdot)}{v_{\sss{\omega}}}		
	\OP{\nmb}
	\nmb^{\sss{\star}},
	\label{eq:TauClosedLoop}	
\end{align}

\end{document}